\documentclass[prd,twocolumn,floatfix,preprintnumbers,letterpaper]{revtex4}
\usepackage{graphicx}
\usepackage{amsmath,amssymb}
\input{epsf}
\usepackage{epsf}

\begin{document}

\def\be{\begin{equation}}
\def\ee{\end{equation}}

\title{The Weak Energy Condition and the Expansion History of the Universe}
\author{A.A. Sen$^1$ and Robert J. Scherrer$^2$}
\affiliation{$^1$Center For Theoretical Physics, Jamia Millia Islamia, New Delhi 110025, India\\
$^2$Department of Physics \& Astronomy, Vanderbilt University,
Nashville, TN 37235, USA}
\date{\today}

\begin{abstract}
We examine flat models containing a dark matter component and
an arbitrary dark energy component, subject only to the constraint
that the dark energy satisfies the weak energy condition.  We determine
the constraints that these conditions place
on the evolution of the Hubble parameter with redshift, $H(z)$,
and on the scaling of the coordinate distance with redshift, $r(z)$.
Observational constraints on $H(z)$
are used to derive an upper bound on the current matter
density.  We demonstrate how the weak energy condition
constrains fitting functions for $r(z)$.

\end{abstract}

\maketitle

\section{Introduction}

Observational evidence \cite{Knop,Riess}
indicates that roughly 70\% of the energy density in the
universe is in the form of an exotic, negative-pressure component,
dubbed dark energy.  (See Ref. \cite{Copeland}
for a recent review).  If $\rho_{DE}$ and $p_{DE}$ are the density
and pressure, respectively, of the dark energy, then
the dark energy can be characterized by the
equation of state parameter $w$, defined by
\begin{equation}
w = p_{DE}/\rho_{DE}.
\end{equation}

Although the simplest possibility for
the dark energy is a cosmological constant,
which has $w=-1$, many other possibilities have
been proposed, including an evolving scalar
field (quintessence) \cite{ratra,turner,caldwelletal,liddle,Stein1},
a scalar field with a non-standard kinetic term
($k$-essence)
\cite{Arm1,Garriga,Chiba1,Arm2,Arm3,Chiba2,Chimento1,Chimento2,Scherrerk},
or simply an arbitrary barotropic fluid with a pre-determined form
for $p(\rho)$, such as the Chaplygin gas and its various
generalizations \cite{Kamenshchik,Bilic,Bento,Dev,Gorini,Bean,Mul,Sen}.

Lacking a definite model for the dark energy (aside from the perennial
favorite cosmological constant), it is interesting to determine
what can be derived from fairly
general assumptions about the nature of the dark energy.
In particular, a plausible assumption about
the dark energy is that it obeys the weak energy condition (WEC).
If $T_{\mu \nu}$ is the energy-momentum tensor of the dark
energy, then the weak energy condition states that
\begin{equation}
T_{\mu \nu} t^\mu t^\nu  \ge 0,
\end{equation}
where $t^{\mu}$ is any timelike vector.  In a
Friedman-Robertson-Walker universe, this reduces to a condition
on the density and pressure:
\begin{equation}
\label{WEC1}
\rho \ge 0,
\end{equation}
and
\begin{equation}
\label{WEC2}
\rho + p \ge 0.
\end{equation}
Although models in which the dark energy violates the WEC are not
inconsistent with current observations
(as first pointed out by Caldwell \cite{Caldwell}), there are
good reasons to believe that the WEC is satisfied
\cite{Carroll,Cline,Hsu1,Hsu2}.

The WEC has already been used previously to constrain
the expansion history of the universe
\cite{Visser,Santos,Perez,Santos2,Santos3,Santos4,Gong}.
These previous studies, however,
all applied the WEC constraints (equations \ref{WEC1} and \ref{WEC2})
to the total cosmological fluid.  Similarly,
Schuecker et al. \cite{Schuecker} applied several
other energy conditions to the total cosmological
fluid.  In this paper, instead, we assume
a model consisting of matter plus a fluid obeying the WEC, and
we determine the corresponding constraints that this places
on the expansion history.

In the next section, we derive the limits that can be placed on
$H(z)$ and $r(z)$ from the WEC.  In Sec. III, we apply these
constraints to analyses of the observations.  Our results
are discussed in Sec. IV.

\section{Consequences of the Weak Energy Condition}

We assume a flat Friedman-Robertson-Walker model, containing a pressureless
matter component and a dark energy component obeying the WEC.  The
radiation contribution to the density at late times is negligible and
can be neglected.  The matter density includes
both baryonic and dark matter, and it scales with redshift $z$
as
\begin{equation}
\rho_M = \rho_{M0} (1+z)^3,
\end{equation}
where the $0$ subscript will refer throughout to present-day values.
The WEC imposes two conditions on the dark energy density:  first,
that
\begin{equation}
\rho_{DE} \ge 0,
\end{equation}
at any redshift (a restatement of equation \ref{WEC1})
and second, that
\begin{equation}
\frac{d\rho_{DE}}{dz} \ge 0, 
\end{equation}
which is a consequence of equation (\ref{WEC2}).

Now consider the consequences of these constraints for the
redshift-dependent
Hubble
parameter $H(z)$, defined by
\begin{equation}
\label{Hdeff}
H(z)^2 = \frac{8 \pi G}{3}(\rho_M + \rho_{DE}).
\end{equation}
It is convenient to work in terms of the present-day value Hubble parameter,
$H_0$, and to use the critical density $\rho_C$ defined
by $H_0^2 = (8 \pi G/3)\rho_C$.  As usual, we take
$\Omega_M = \rho_{M0}/\rho_C$ and $\Omega_{DE}  = \rho_{DE0}/\rho_C$,
and our assumption of a flat universe gives $\Omega_M + \Omega_{DE} = 1$.
Our $\Omega$'s will always refer to present-day quantities; we
omit the $0$ subscript in this case for simplicity.  Then equation
(\ref{Hdeff}) can be rewritten as
\begin{equation}
\label{Htemp}
\widetilde H(z)^2 = \Omega_M(1+z)^3 + (1-\Omega_M) (\rho_{DE}(z)/\rho_{DE0}),
\end{equation}
where we have defined $\widetilde H(z) \equiv H(z)/H_0$.
The WEC forces $(\rho_{DE}(z)/\rho_{DE0}) \ge 1$ for $z > 0$, so
equation (\ref{Htemp}) becomes
\begin{equation}
\label{H1}
\widetilde H(z)^2 \ge \Omega_M(1+z)^3 + 1 - \Omega_M.
\end{equation}
Going back to equation (\ref{Htemp}) and taking the derivative
with respect to $z$ (which we denote throughout
with a prime) gives:
\begin{equation}
2 \widetilde H(z) \widetilde H^\prime(z) = 3 \Omega_M (1+z)^2
+ [(1-\Omega_M)/\rho_{DE0}] \frac{d \rho_{DE}}{dz}.
\end{equation}
Now the WEC implies that $d\rho_{DE}/dz \ge 0$, so we get
\begin{equation}
\label{H2}
\widetilde H(z) \widetilde H^\prime(z) \ge \frac{3}{2} \Omega_M (1+z)^2.
\end{equation}
Equations (\ref{H1}) and (\ref{H2}) give the constraints that
the WEC for the dark energy
places on the redshift-dependent Hubble parameter.  Note
that these constraints are not independent.  The limit in equation
(\ref{H2}), together with $\widetilde H(0) = 1$, implies the limit
given in equation (\ref{H1}), but the converse is not true.
Although equation (\ref{H2}) gives the stronger limit,
we include both limits because current data can provide some estimates
for $\widetilde H(z)$ (as in the next section), but are too poor to
provide any limits on $\widetilde H^\prime (z)$.  By construction,
the standard $\Lambda$CDM model saturates both limits.
Equation (\ref{H2}) was previously introduced by Sahni and Starobinsky
in the context of quintessence models \cite{SS} and by Boisseau et al.
in the examination of scalar-tensor models \cite{Bois}.

Now consider the coordinate distance $r(z)$, defined by
\begin{equation}
\label{rz}
r(z) = \int_0^z \frac{dz^\prime}{H(z^\prime)}.
\end{equation}
The coordinate distance is important because it is directly
related to the luminosity distance, $d_L$, through
$d_L = c(1+z)r(z)/H_0$, and it is $d_L$ which is measured in
supernova redshift surveys.  Hence, a considerable  effort has been
put into designing parametrizations for $r(z)$ to fit to the supernova
data \cite{Huterer,Chiba3,Saini,Efsth,Weller,Pad1,Pad2,Li}.  The purpose
of many of these investigations is to go from a best-fit form
for $r(z)$ to the potential for an underlying quintessence
model.

The derivative of equation (\ref{rz}) gives
\begin{equation}
r^\prime(z) = \frac{1}{H(z)},
\end{equation}
which, when combined with equation (\ref{H1}), yields
\begin{equation}
\label{r1}
r^\prime(z) \le H_0^{-1} [\Omega_M(1+z)^3 + (1-\Omega_M)]^{-1/2}.
\end{equation}
Similarly, equation (\ref{H2}) yields
\begin{equation}
\label{r2}
- \frac{r^{\prime \prime}(z)}{{r^\prime}(z)^3} \ge \frac{3}{2} H_0^2 \Omega_M (1+z)^2.
\end{equation}
Equations (\ref{r1}) and (\ref{r2}) give the dark-energy
WEC constraint
on the coordinate distance.  As for the limits on $\widetilde H(z)$,
equation (\ref{r2}) implies (\ref{r1}),
but the converse is not true.

\section{Comparison with Observations}
\begin{figure}[t]
\centerline{\epsfxsize=3.3truein\epsffile{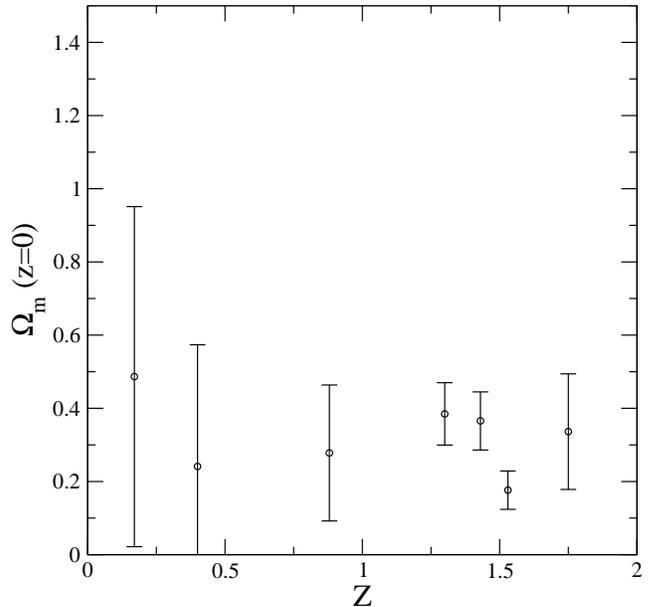}}
\caption{Upper bound on the present-day value of $\Omega_M$ from the weak energy condition
applied to the dark energy, using the $H(z)$ values in
Ref. \cite{Simon}, for $H_0 = 73 \pm 3$ km s$^{-1}$ Mpc$^{-1}$.}
\end{figure}

Consider first our limits on the evolution of $H(z)$.  Estimates of
$H(z)$ were derived by Simon, Verde, and Jimenez \cite{Simon} using
passively evolving galaxies; these limits have been
used to constrain cosmological parameters in dark energy
models \cite{Ratra}.  A second approach to deriving $H(z)$ based
on the Supernova data has been explored in Refs. \cite{Daly}
and \cite{Shafieloo}.  We will use the results of \cite{Simon}
in our discussion here.

Equation (\ref{H1}) can be rewritten as
\begin{equation}
\Omega_M \le \frac{\widetilde H(z)^2 - 1}{(1+z)^3 - 1}.
\end{equation}
Thus, a single value of $\widetilde H(z)$ can provide an upper
bound on $\Omega_M$.  Since our bound applies to any dark
energy model satisfying the WEC,
we do not attempt to fit any particular model (as was done
in Ref. \cite{Ratra}); rather, we calculate the upper bound
individually for each $H(z)$ measurement.  The upper bound
on $\Omega_M$ depends on
$\widetilde H(z) = H(z)/H_0$, so our results
will naturally be sensitive to the value of $H_0$.  Following Ref.
\cite{Ratra} we consider two priors for $H_0$:  $H_0 = 73 \pm 3$
km s$^{-1}$ Mpc$^{-1}$ from WMAP \cite{Spergel},
and $H_0 = 68 \pm 4$ km s$^{-1}$ Mpc$^{-1}$ from the median
statistics analysis in Ref. \cite{Gott}. Our results
are shown in Figs. 1 and 2 (where all error bars are 1-sigma).
For $z=1.53$, where the error bars are smallest,
we get the tightest upper bound on
$\Omega_M$:  $\Omega_M \le 0.18 \pm0.05$ for $H_{0} =  73 \pm 3$
km s$^{-1}$ Mpc$^{-1}$,
and $\Omega_M \le 0.21 \pm 0.06$
for $H_{0} = 68 \pm 4$ km s$^{-1}$ Mpc$^{-1}$.  (For a very different
approach, see Ref. \cite{Zhang}).
\begin{figure}[t]
\centerline{\epsfxsize=3.3truein\epsffile{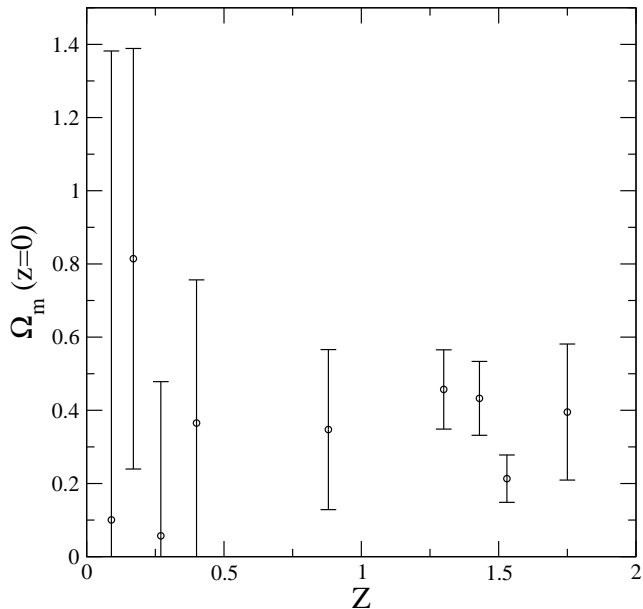}}
\caption{As Fig. 1, for
$H_0 = 68 \pm 4$ km s$^{-1}$ Mpc$^{-1}$.}
\end{figure}

Next we consider the consequences of the limits on the coordinate
distance $r(z)$ given by equations (\ref{r1}) and (\ref{r2}).
The quantity observers actually measure is the apparent magnitude
$m(z)$, given by
\begin{equation}
m(z) = {\cal{M}} + 5 Log_{10} (D_{L}(z)),
\end{equation}
where $D_L$ is the Hubble-free luminosity distance,
\begin{equation}
D_{L}(z) = (1+z) \int_{0}^{z} d{z^\prime}{H_{0}\over{H({z^\prime})}},
\end{equation}
and
$\cal{M}$ is the magnitude zero-point offset,
which depends on the absolute magnitude $M$ as
\begin{equation}
{\cal{M}} = M + 5Log_{10}\left({H_{0}^{-1}\over{Mpc}}\right) + 25 = M - 5Log_{10}h + 42.38.
\end{equation}
The distance modulus given by the SNIa data is defined as
\begin{equation}
\mu(z) = m(z) - M = 5Log_{10}(D_{L}(z))+\mu_{0},
\end{equation}
where $\mu_{0} = 42.38 - 5Log_{10}h$.
\begin{figure}[t]
\centerline{\epsfxsize=3.5truein\epsffile{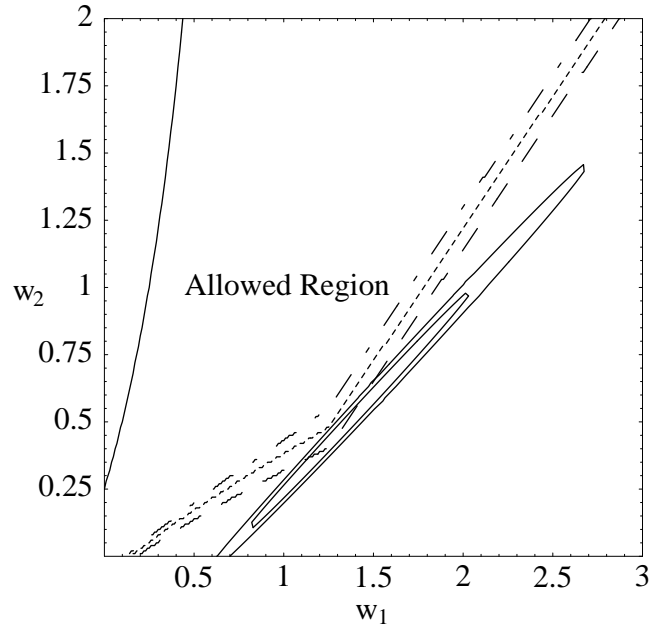}}
\caption{The region in the $w_1$, $w_2$ plane allowed by the weak energy condition
for the parametrization
of $r(z)$ given in equation (\ref{rzfit}).
The region between the two sets of curves is allowed.  Right-hand boundaries are given
for $\Omega_M = 0.2$ (dashed),
$\Omega_M = 0.3$ (dotted) and $\Omega_M = 0.4$
(dot-dash).  Left-hand boundary (solid)
is independent of $\Omega_M$.
The two ellipses are the
$1\sigma$ and $2\sigma$ contours obtained by fitting equation
(\ref{mufit}) to the supernova data without the weak energy constraint.}
\end{figure}

As an example, we now choose a representative fitting function, translate it
into a fitting function for $r(z)$, and then
determine how the WEC constrains the parameters of the fitting function.
In this paper,
we use the fitting function for $\mu$, first introduced by
Padmanabhan and Choudhury \cite{Pad2}, given by
\begin{equation}
\label{mufit}
\mu_{fit} = \mu_{0} + 5Log_{10}\left[{z(1+w_{1}z)\over{(1+w_{2}z)}}\right],
\end{equation}
where $\mu_{0}, w_{1}$ and $w_{2}$ are the three independent fitting parameters.
By comparing equations (21) and (22), we can write
\begin{equation}
D_{L} = {z (1+w_{1} z)\over{(1+w_{2}z)}},
\end{equation}
so that
\begin{equation} 
\label{rzfit}
 r(z) = {1\over{H_{0}}}{z(1+w_{1}z)\over{(1+z)(1+w_{2}z)}}.
\end{equation}
Since we have an analytic form
for $r(z)$, we use equation (\ref{r2}) alone to determine
the values of $w_1$ and $w_2$ which violate the WEC; any
$r(z)$ which satisfies equation (\ref{r2}) will automatically
satisfy equation (\ref{r1}).  We require equation (\ref{r2})
to be satisfied for $z < 2$, the range over which the supernova
data extend.

The allowed
region for $w_1$ and $w_2$ is displayed in Fig. 3.  As expected,
the constraints are tighter for larger values of $\Omega_M$,
but rather surprisingly, the excluded region
in parameter space is rather insensitive to the assumed value of $\Omega_M$
(and a significant region of parameter space is excluded even in the
limit $\Omega_M \rightarrow 0$).
Also in Fig. 3, we display the confidence regions for $w_1$ and $w_2$
from the supernova data, with no assumptions about the equation of state
for the dark energy.
We use 60 Essence supernovae \cite{Essence},
57 SNLS supernovae \cite{SNLS} and 45 nearby supernovae.
We have also included the new data release of 30 SNe Ia detected by
HST and classified as the Gold sample by Riess et al. \cite{Riess1}.
The combined data set can be found in Ref. \cite{davis}.
Clearly, the best-fit values for $w_1$ and $w_2$ lie slightly outside
of the allowed region for $\Omega_M \ge 0.3$.  Not too much should be read into this:  given
that $\Lambda$CDM models provide a good fit to all current observations,
and such models saturate our bounds, we would expect the best fit
parameters to lie near the boundary of the excluded region.  The important
point is that Fig. 3 shows how one can use our constraints on $r(z)$
to eliminate regions of parameter space for which the dark energy
violates the WEC.

Of course, it is also possible to parametrize $w$ as a
function of $z$ and use this parametrization to derive a
form for $r(z)$ (see, e.g., Ref. \cite{Efsth}); in this
case the WEC is trivially satisfied as long as the parametrization
for $w(z)$ forces $1+w(z)$ to be nonnegative.

\section{Discussion}

We have examined how the WEC constrains the redshift evolution of both 
the Hubble parameter $H(z)$ and coordinate distance
$r(z)$.  The constraint on $H(z)$ can be combined with observations
of $H(z)$ to put upper bounds on $\Omega_M$.
While the scatter in these estimated upper limits is large,
as are the errors, the important point is that these are
generic upper limits, independent of the nature of the dark energy (as
long as it satisfies the weak energy condition).
Improved measurements, particularly of $H(z)$, will strongly
improve this upper bound.
The constraints on $r(z)$ do not provide similarly useful limits,
but they can be applied to any parametrization of $r(z)$ to eliminate
in advance any regions of parameter space in which the dark energy
violates the weak energy condition.

What happens if the WEC is violated by the dark energy?  If one
allows for arbitrary evolution, then there are clearly no
constraints on $r(z)$ and $H(z)$.  An intermediate case, which provides
weaker limits than the ones we have discussed, is when the WEC applies
to the total fluid (matter and dark energy together)
\cite{Visser,Santos,Perez,Santos2,Santos3,Santos4,Gong}.  In this
case, for example, there is no limit corresponding to equation
(\ref{H1}), while the limit corresponding to equation (\ref{H2})
becomes $\widetilde H(z) \widetilde H^\prime(z) \ge 0$.
Thus, this version of the WEC provides no bound on $\Omega_M$,
although it does constrain the evolution of $H(z)$.
Applying the WEC to the total fluid
also yields constraints on $r(z)$;
these are discussed in detail in Ref. \cite{Santos}.  These
constraints are weaker but more general than the constraints
we have obtained by applying the WEC to the dark energy alone.

\acknowledgements 

R.J.S. was supported in part by the Department of Energy (DE-FG05-85ER40226).
We thank D. Polarski for helpful comments.

\end{document}